# Dynamic manipulation of mechanical resonators in the high amplitude regime through optical backaction

Mahmood Bagheri, Menno Poot, Mo Li, Wolfram P. H. Pernice, Hong X. Tang

*Department of Electrical Engineering, Yale University, New Haven, CT 06511*

**Cavity optomechanics enables active manipulation of mechanical resonators through backaction cooling and amplification[1,2]. This ability to control mechanical motion with retarded optical forces has recently spurred a race towards realizing a mechanical resonator in its quantum ground state[3,4,5,6,7,8,9]. Here, instead of quenching optomechanical motion, we demonstrate high amplitude operation of nanomechanical resonators by utilizing a highly efficient phonon generation process. In this regime, the nanomechanical resonators gain sufficient energy from the optical field to overcome the large energy barrier of a double well potential, leading to nanomechanical slow-down and zero frequency singularity, as predicted by early theories[10]. Besides fundamental studies and interests in parametric amplification of small forces[11], optomechanical backaction is also projected to open new windows for studying discrete mechanical states[12,13] and to foster applications[14,15,16,17]. Here we realize a non-volatile mechanical memory element, in which bits are written and reset via optomechanical backaction by controlling the mechanical damping across the barrier. Our study casts a new perspective on the energy dynamics in coupled mechanical resonator – cavity systems and enables novel functional devices that utilize the principles of cavity optomechanics.**

Classical and quantum dynamics of nanomechanical systems promise new applications in nanotechnology[18,19] and fundamental tests of quantum mechanics in mesoscopic objects[2,9]. Recent development of nanoscale electromechanical (NEMS) and optomechanical systems has enabled cooling of mechanical systems to their quantum ground state[7,8], which brings the possibility of quantum information processing with mechanical devices[20,21]. On the other hand, for practical application at room temperature — such as signal processing[22] and mass/force sensing[23] — it is desirable to operate nanomechanical devices at high amplitudes. However, scaling a mechanical resonator to nanoscale dimensions reduces its dynamic range dramatically[24]: when a coherent drive of increasing strength is applied to the nanomechanical resonator, Duffing nonlinearities shift the resonance frequency away from the drive and the amplitude hardly increases. Therefore, reaching high amplitude in a NEMS using coherent driving is difficult. Here we show that by exploiting cavity optomechanics these limitations of traditionally operated NEMS resonators can be overcome.

Coupling a mechanical resonator to an optical cavity leads to a variety of effects such as optical bistability[25], the optical spring effect[26], motion-induced transparency[27,28,29], and damping and amplification of the resonator's thermal motion. It is the latter effect that enables us to operate our devices in the high-amplitude regime, where the nanomechanical resonator can overcome the large energy barrier that separates its two stable positions. This enables our observation of the long-sought after zero-frequency anomaly in a nanomechanical system[10]. Furthermore, by utilizing real-time optical cooling, we controllably quench the oscillator into one of its potential-well minima from the high-amplitude state. This way, we demonstrate all-optical operation of a mechanical memory element which stores information in the two mechanically stable states of the resonators. Such memories, described by a canonical double-well potential with a large central energy barrier, require large vibration amplitudes to switch between discrete, mechanical states[30]. Since the barrier is sufficiently larger than the thermal energy, our mechanical memory is non-volatile and immune to electromagnetic perturbation, environmental fluctuations and leakage[31], and presents itself as a viable building block for the current effort in developing mechanical computing engines[32,33,34,35].

A rendering of the optomechanical system used in this study is shown in Fig. 1a. It consists of a nanometre-sized flexural resonator integrated in an optical race-track cavity, fabricated from commercially available silicon-on-insulator (SOI) wafers. The nanomechanical resonators have dimensions of 10 µm × 500 nm × 110 nm, an effective mass of $m_{eff}$ = 1.0 pg, and a fundamental resonance frequency of $\Omega_m/2\pi$ ~ 8 MHz with mechanical damping rate of $\Gamma_0/2\pi$ = 2.1 kHz. The race-track cavity has a free spectral range of 2 nm and a typical linewidth of 10 GHz. Optomechanical interactions are mediated by optical gradient forces between the nanomechanical resonator and the adjacent substrate[36]. The mechanical resonator is separated from the substrate by about 250 nm to have large optomechanical coupling rates ($g/2\pi$ = 1 GHz/nm where $g = d\omega_c/dx$, see Supp. Mat.) and yet to allow large oscillation amplitudes. Due to the residual compressive stress introduced by the SOI wafer bonding process, the free-standing doubly clamped beams are slightly buckled and have two stable configurations at rest[37]: buckled up and buckled down (see Fig. 1b). Therefore, the out-of-plane motion of the buckled beam can be described by a double-well potential, where both the 'up' and 'down' states correspond to the minima in the potential. The thermomechanical displacement noise spectra in Fig. 1e show that the two states have slightly different mechanical resonance frequencies, which indicates that the double-well potential is not completely symmetric.

The two mechanical states are discriminated in optical transmission measurements because the optical mode has a different effective refractive index in the two states: when the waveguide is closer to the substrate (buckled-down state) the effective refractive index is larger than in the buckled-up state as the optical mode interacts stronger with the substrate. Therefore the optical cavity resonance shifts towards longer wavelengths when the resonator flips from the buckled-up state to the buckled-down state. Consequently, the optical cavity has distinct optical resonances in the two stable configurations as shown in low power optical transmission spectra in Fig. 1c. The low power spectrum probes the static optomechanical resonances, but at high power the mechanical resonator starts to oscillate when the pump wavelength is scanned close to the optical resonance. Fig. 1d shows the optical transmission spectrum measured when the input optical power is well above the threshold for self-sustained oscillations (SSO) of ~600 µW. When the wavelength is scanned across the optical resonance, the transmission no longer shows the low-power Lorentzian shape, rather the resonance is dragged from the "up" to the "down" state: as soon the laser is blue detuned w.r.t. the "up" state, the self-sustained oscillations start and the cavity frequency oscillates back and forth with an amplitude $A_{p-p} \cdot g$, where $A_{p-p}$ is the mechanical resonator oscillation amplitude. The SSO in turn modulate the optical transmission at the mechanical oscillation frequency, indicated by the high-frequency components of the optical transmission (Fig. 1d). The SSO appear in the entire wavelength range between the "up" and "down" state optical modes, which indicates that the energy of the resonator exceeds the energy barrier between the two states. Stable self-sustained oscillators are highly desirable as they can provide low-noise high-frequency signals for potential applications in precision on-chip timekeeping, communication technology, and sensing[38]. Generally, to reduce the effect of noise in such applications, it is essential to increase the oscillation amplitude as much as possible by pumping a large number of phonons into the oscillator.

The high amplitude optomechanical system we study here has very distinctive nonlinear dynamics in both the mechanical and optical domain. In the conventional framework of cavity optomechanics [1,2](see Fig. 2a), a blue detuned input pump laser (at frequency $\omega_p$) resonantly enhances the Stokes sideband, while the off-resonant anti-Stokes sideband is suppressed. It is therefore more likely for a pump photon in the cavity to emit a phonon than to absorb one, thereby amplifying the motion. This dynamical backaction of the cavity on the resonator can be represented by a negative damping rate $\Gamma_{BA}$ which reduces the total mechanical damping rate $\Gamma$ from its intrinsic value $\Gamma_0$ to $\Gamma_0 + \Gamma_{BA}$. On the other hand, for a red-detuned pump, $\Gamma_{BA}$ is positive and leads to cooling. When the blue-detuned pump power is large enough $\Gamma_{BA}$ can become so negative that the total damping rate of the mechanical resonator $\Gamma_0 + \Gamma_{BA}$ vanishes, leading to regenerative oscillations sustained by the static pump power.

In cavity optomechanics, the resolved sideband regime (RSR) is regarded as the most efficient way to cool the mechanical resonator. One might expect that the same is true for amplification. This is, however, untrue, since in the RSR only a single sideband is available, and one pump photon can only emit a single phonon[39], as illustrated in Fig. 2b. The unresolved sideband regime (USR) is more preferable for efficient phonon generation. In this case many sidebands (Fig. 2c) are available for phonon emission and absorption and can even interfere with each other[40]. Consequently, the net effect is that a single blue-detuned photon can emit many phonons[41]. (A detailed analysis of phonon generation process in USR can be found in the Supp. Mat.) In a system operated in the USR, the oscillations can actually shift the cavity resonance significantly beyond the cavity linewidth (i.e. $A_{p\text{-}p}\, g \gg \gamma$), as shown in Fig. 2d. As the oscillations move the cavity out of resonance from the input pump light, the cavity is empty most of the time (see back plot of Fig. 2d) and the resonator undergoes damped harmonic motion. Only when the cavity becomes resonant with the pump, it fills quickly with photons that rapidly emit (or absorb) a large amount of phonons in an avalanche process, exciting SSO to large amplitudes. The red regions in the color plot in Fig. 2d indicate when the resonator gains energy, whereas in the blue regions it loses energy. In this work, we are able to experimentally excite the nanomechanical resonator to such high amplitudes by using a deeply unsolved–sideband cavity system ($\gamma/2\pi = 10$ GHz vs. $\Omega_m/2\pi \sim 8$ MHz) in combination with a large cavity dynamic range[42] ($\gamma/g = 10$ nm), so that very large oscillation amplitudes (>300nm, equivalent to more than $10^{12}$ phonons) are obtained. This oscillation amplitude exceeds the Duffing limit by almost two orders of magnitude (See Supp. Mat. for the measurement of Duffing limits).

We illustrate the high amplitude optomechanical amplification processes in the time domain (see Supp. Mat. for experimental details). When the nanomechanical resonator resides in the buckled-down state, an optical drive blue-detuned from the optical cavity resonance initiates the optomechanical amplification (phonon generation) process and starts coherent oscillations of the nanomechanical resonator. When the kinetic energy of the mechanical resonator becomes larger than the double-well potential barrier, the beam can relax into either of the two potential wells after the excitation has been turned off[43]. Fig. 3a shows a schematic representation of the amplification process, while Figs. 3b&c illustrate the situations when the nanomechanical resonator freely decays into either of the stable states.

Fig. 3d shows the time evolution of the nanomechanical resonator response when the beam is initially in its buckled-down state. When the blue-detuned laser is switched on at zero time, the resonator starts to oscillate coherently. The amplitude grows in time until it saturates after about 250 μs. The resonator then oscillates in a high phonon-number state above the double-well potential barrier. Note, that for A > ~ γ/g the cavity is no longer a linear transducer and the output signal has a distorted waveform (see Figs. S6 and S7b), which makes direct determination of the amplitude difficult. Here instead, we monitor the mechanical resonant frequency to map out the vibration amplitude: in a double-well potential as the oscillation amplitude goes up and down, the resonant frequency varies over a large range – a feature that is not available in a linear resonator. Figure 3g shows the digital Fourier transform spectra of the nanomechanical response calculated for different times as the coherent oscillations build up. Initially, the beam is in the "down" state and oscillates with its intrinsic resonance frequency (7.4 MHz). However, when it is excited to high-amplitudes and its energy exceeds the potential barrier, the oscillator frequency has decreased to 3.1 MHz. Such a drastic change in the resonance frequency is characteristic of a mechanical resonator with a double-well potential, and is different from those induced by the optical spring, i.e. the effect that the photon pressure is displacement dependent, as reported in other optomechanical systems[26].

This resonant behaviour is further confirmed by switching off the pump laser and subsequently monitoring the nanomechanical motion as its energy is dissipated at a rate $\Gamma_0$ while the resonator relaxes towards one of its equilibrium points (Fig. 3e and f). This is done using a weak probe laser. Since the excitation pump is off and only the weak optical probe samples the response of the nanomechanical resonator, the optical backaction does not contribute to the frequency change of the resonator (Fig. 3h and i), confirming the mechanical origin of the observed frequency shift. Without active control the resonator can relax into either the buckled-down or buckled-up states, as shown in the middle and right panels respectively. Later, we show that by manipulating the optomechanical damping $\Gamma_{BA}$ (i.e. by actively cooling) we can reproducibly control in which well the resonator relaxes.

Figure 4a shows the continuous evolution of the oscillation frequency as it relaxes from the high energy state towards the two potential minima. The solid curves in Fig. 4a represent the instantaneous frequency of the nanomechanical beam obtained from the time traces, whereas the markers show the resonance frequency of the mechanical beam obtained from the digital Fourier transform spectra at different times. The low energy frequencies closely match the resonance frequencies obtained from the displacement noise spectra (Fig. 1e). The asymmetry in the potential implies a bifurcation once the beam gets trapped in either of the two wells during the relaxation from a high-amplitude state. Figure 4a shows that the bifurcation point coincides with a divergence in the oscillation period. This is the first demonstration of the long-sought-after zero frequency singularity in nanomechanical devices, which can be explained as follows: when the resonator energy approaches the top of the potential barrier, the beam's kinetic energy vanishes, so the resonator slows down. The resonator thus dwells a long time near this potential extremum, before speeding up again after getting trapped in either of the two potential wells[44] as illustrated in Fig. 4c. In the actual measurement, the lowest frequency achieved depends statistically on the point of the phase trajectory at which the excitation is turned off as well as the force noise experienced by the resonator during its ring-down (see Supp. Mat.). This operation regime of NEMS provides a new mechanism to tune the resonance frequency of the mechanical beam over a large range. This wide tuneability of resonant frequency has important consequences for building tuneable NEMS oscillators and for synchronization of large NEMS oscillator arrays[45].

The dependence of the mechanical resonance frequency on the beam energy portrays the trajectory of the resonator and the slowed resonator effect when it is traversing the double-well potential. This dependence is used to gauge the high amplitude motion of the resonator when the mechanical beam is oscillating between the two buckled states.

(This would be otherwise not possible due to the limited sensing range of optical cavity as explained earlier). The instantaneous frequency (i.e. the inverse of the duration of each period) is extracted from the measured time traces. Then, using the relation between the motion amplitude and the oscillation period, the amplitude is determined. Figure 4b shows the extracted peak-to-peak oscillation amplitude $A_{p-p}$ as the mechanical resonator relaxes towards its resting points. The novel operation of the optomechanical system thus yields amplitudes as high as $A_{p-p}$ = 350 nm. As shown in Fig. 4c, this corresponds to a resonator energy of 9 fJ, or equivalently, to $4 \times 10^{12}$ phonons in the resonator.

It is notable that once the beam is elevated above the potential barrier and the excitation laser is turned off, the state into which the resonator relaxes into is undetermined. However, controlling the final state is an essential ingredient for applications. We show that a deterministic selection of the final state is possible by utilizing optomechanical cooling in a two-color scheme. In this scheme, first the amplification pump laser is tuned to excite the coherent SSO. Next, the cooling laser is switched on to damp the oscillations towards either of the two mechanical states. The final state is selected by tuning the damping laser to the red side of the optical resonances of the selected state (Fig. 2c): When applying the cooling laser between the two optical resonances, it damps the motion in the 'up' potential well but amplifies the motion in the "down" potential well, therefore a larger probability of the beam ending up in the "up" state can be expected. On the other hand, when the cooling laser is red-detuned w.r.t. the down-state optical mode, the cooling power into the down state is much larger than into the up state, which is many linewidths away, so that the "down" state can be selected as the final state. By repeatedly exciting the resonator to large oscillations and subsequently cooling its motion, the asymmetry in the final state probability can be mapped out.

Figure 5a shows the probability of cooling into the 'up' and 'down' state as a function of detuning. There is a clear wavelength-dependent asymmetry in the probability: around $\lambda_1$ = 1560.75 nm the resonator always relaxes into the "up" state, whereas at $\lambda_0$ = 1561.00 nm it is always projected into the 'down' state. This enables the operation of our optomechanical system as a non-volatile memory element where the 'up' and 'down' buckled states of the beam correspond to the '1's and '0's. The barrier between the states is large compared to thermal energy (5 fJ corresponds to 350,000,000 K), so after the data is stored, the memory state is maintained without requiring optical power to retain it. This is in contrast to bit storage in the phase of an excited mechanical resonator[32,33,46] which loses stored information when the input power is disconnected. Figure 5b shows a schematic representation of the sequence that we employ to write, read and erase information. In one operation cycle, first a reset pulse of blue detuned light is used to raise the resonator energy over the barrier via self oscillations. Then the data is written onto the resonator by damping the motion into either of the two states with a writing pulse of light red-detuned at selected wavelength (either $\lambda_1$ or $\lambda_0$). In the last part of the cycle, a weak probe pulse tuned to one of the optical resonances associated with the buckled-up or buckled-down states is used to read the state of the memory.

Figure 5c shows the experimental realization of the bit operation sequence described above. A series of consecutive '1's and '0's are written into the memory element. Each operation consists of an amplifying signal (blue pulses) that starts the coherent oscillations of the nanomechanical resonator followed by a cooling signal that selectively dampens the nanomechanical resonator into buckled-up (pink) or buckled-down (red) states. Sequences of 010101 and 001001 are repeated thousands of times with no error detected in the readout signal. The device shows no signs of wear or fatigue after more than one billion operations. By adding the laser power times the duration of the excitation and cooling pulses the energy cost of the data operations can be estimated. The current value of 10 µJ can be reduced by using optical cavities with higher quality factors ($Q_{opt}$ = 20,000 for the current devices) and by engineering the elastic energy profile of nanomechanical resonators.

In summary, we find by leveraging cavity optomechanical backaction in the unresolved sideband regime, nanomechanical resonators are operable at unprecedented high amplitudes. The dynamic control offered by optical backaction presented here enables the observation of the zero-frequency anomaly, active manipulation of mechanical states of a beam resonator, and modification of the damping dynamics in real time through external optical signals. As the amplitude is tuned up and down the potential well, the resonance frequency varies over a wide range. A non-volatile optical memory that switches between the binary states based on optical cooling and amplification is demonstrated. The memory element retains its data after it is exposed to environmental variations over the measurement period.

Our scheme further brings on-chip optical cooling and amplification to a practical device application. This new type of nanomechanical memory element is built on a scalable silicon platform and may be exploited for optically multiplexed memory storage, addressable mechanical switch array, tuneable optical filters, and reconfigurable optical networks.

**Methods**

**Device Fabrication.** The devices were fabricated from silicon-on-insulator wafers with a 110-nm thick silicon layer and a 3 µm thick layer of buried oxide. Nano-photonic circuits were patterned by electron-beam lithography using a Vistec Ebpg 5000+ Lithography system, and a plasma dry etch using an inductively-coupled $Cl_2$ gas chemistry. A wet etch release window was patterned using direct write photo-lithography, followed by a buffered oxide etch. Finally, the beams were dried in a critical point dryer. Nanomechanical resonators with lengths between 10 µm and 20 µm in 2.5-µm steps were fabricated. All device lengths show two stable states at rest; the measurements presented here were measured on a 10 µm long beam.

**Measurement.** The measurements were performed in a vacuum chamber with a base pressure of 100 µTorr. Light was coupled in and out of the bus waveguide using an array of cleaved single-mode polarization-maintaining fibers aligned to the input/output grating couplers (Fig. 2a). The typical coupling efficiency was 30% per grating coupler. Three tunable diode lasers (two Santec TSL-210 and one HP 8648F) were used to amplify the motion of the nanomechanical beam and subsequently quench the motion into the target state (see Supp. Mat. for more details)

**Acknowledgements** The authors thank Dr. M. Rooks of the Yale Institute for Nanoscience and Quantum Engineering for helping with the ebeam lithography, and Michael Power for helping with device fabrication. We acknowledge funding support from DARPA/MTO ORCHID program through a grant from AFOSR. H.X.T. acknowledges support from a Packard Fellowship in Science and Engineering and a career award from National Science Foundation. M.P acknowledges a Rubicon fellowship from the Netherlands Organization for Scientific Research (NWO) / Marie Curie Cofund Action.

**Competing Interests** The authors declare that they have no competing financial interests.

**Correspondence** Correspondence and requests for materials should be addressed to H. X. Tang. (e-mail: hong.tang@yale.edu).

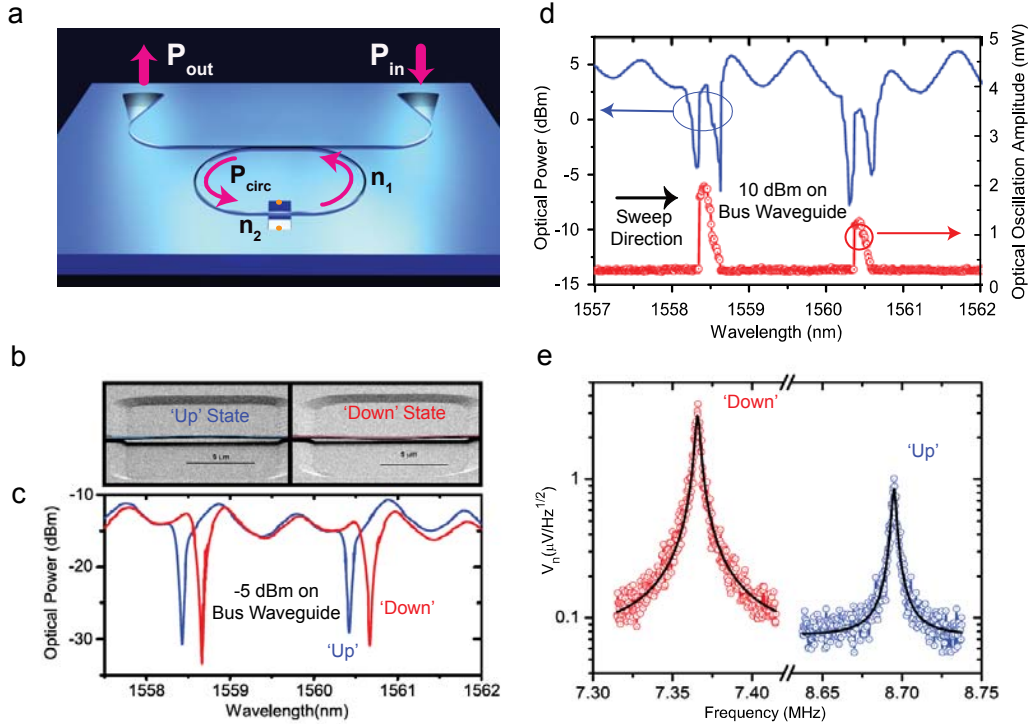

**Figure 1 Nanomechanical beam in a double-well potential.** (a) Schematic representation of the nanomechanical beam embedded in a photonic race-track cavity. Light is sent into the bus waveguide via grating couplers (triangles). By underetching a part of the cavity a flexural resonator is formed. (b) Scanning electron micrographs of the nanomechanical beam in its buckled-up (left) and buckled-down (right) states. (c) Optical transmission spectra of the photonic circuit measured at low input power when the nanomechanical beam is in the buckled-up (blue curve) and buckled-down (red curve) states. (d) Optical transmission spectrum of the race-track cavity measured at high input power. Blue trace: dc transmission; red trace: ac oscillation amplitude. (e) Thermo-mechanical noise spectra measured in the buckled-up (blue curve) and buckled-down (red curve) state. The solid lines are harmonic oscillator responses fitted to the data (symbols).

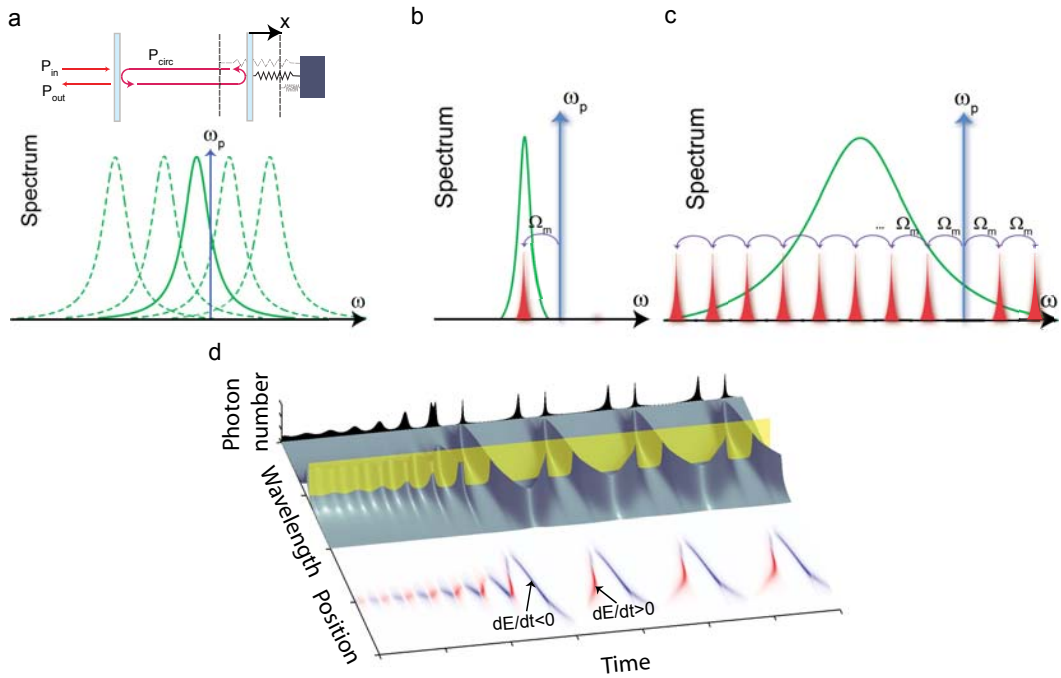

**Figure 2 Large oscillations in an optomechanical system.** (a) Schematic representation of large oscillations in an optomechanical system. The cavity is illuminated with a pump laser with frequency $\omega_p$ and power $P_{in}$. The line shape of the cavity for an undisplaced resonator is indicated by the solid green line; a deflection of the resonator shifts the cavity frequency (dashed lines). (b, c) Phonon emission in the resolved sideband regime (b) and in the unresolved sideband regime (c). In the RSR, only one sideband is present, whereas many phonons can be emitted in the USR. (d) Simulated ring-up of the SSO induced by a blue detuned pump. The yellow plane represents the pump laser wavelength, which intersects the moving cavity resonance. The color plot shows the energy gain: Red indicates regions where the resonator gains energy, whereas times at which the resonator looses energy (either due to damping or to the cavity) are shown in blue. The back panel shows the time-dependence of the cavity occupation.

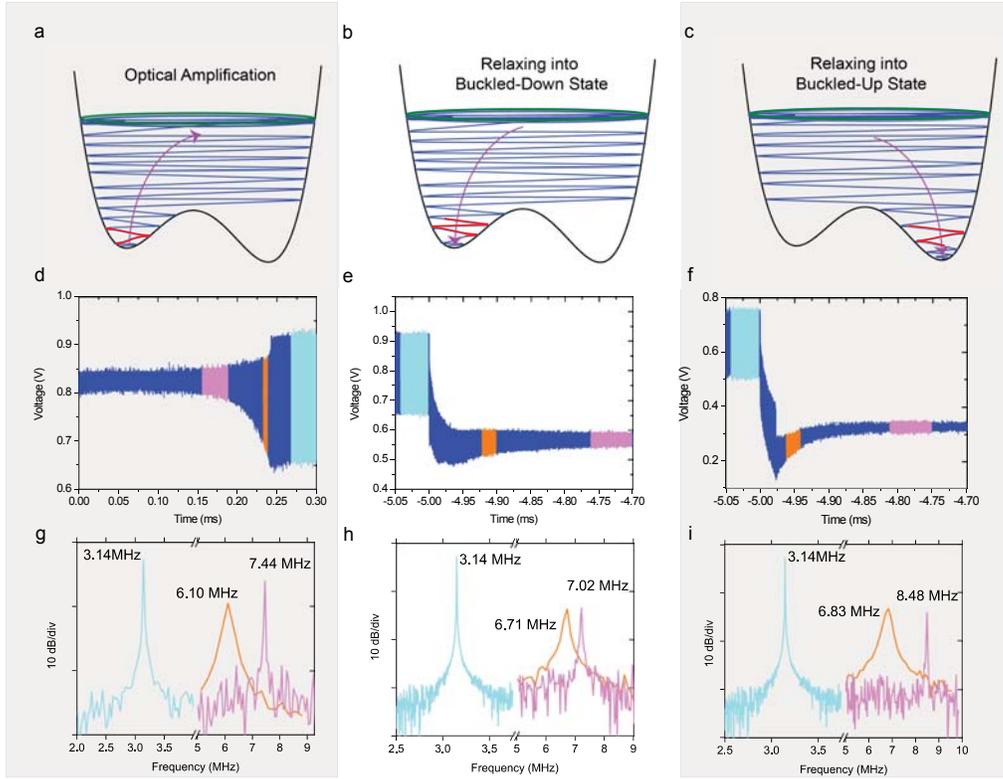

**Figure 3 Optomechanical amplification and cooling of the nanomechanical beam in a double-well potential.** Ring-up to the high amplitude state (left panels), and free decay of the nanomechanical resonator into the buckled-down (middle panels) and buckled-up states (right panels). (a)-(c) schematic of the processes. (d)-(f) measured optical transmission displaying the temporal response of the nanomechanical beam. (g)-(i) Fourier spectra of different parts of the time traces shown in (d)-(f) respectively. Note, that the probe wavelength in panels (f) and (i) was adjusted from that in the other panels in order to have a good transduction in the up state.

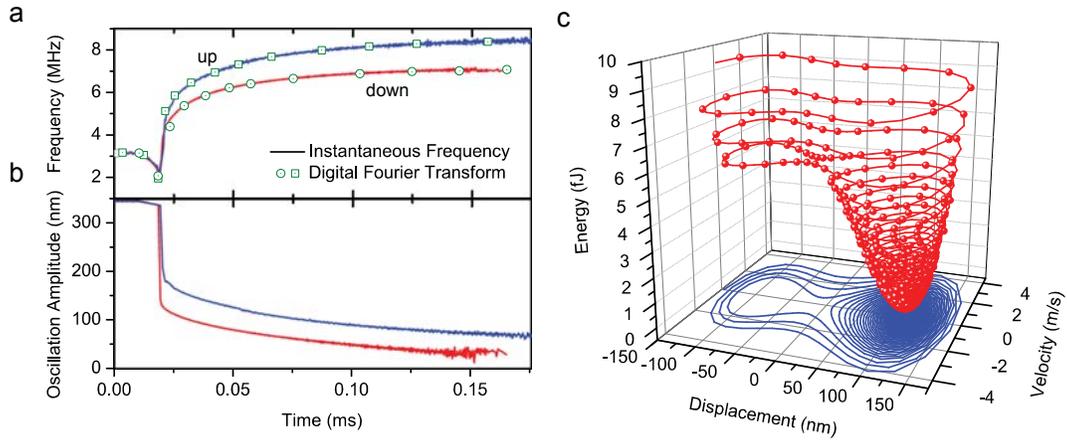

**Figure 4 Ring-down and zero-frequency anomaly**. Measured time-dependence of the oscillation frequency (a) and the oscillation amplitude (b) during free decay into the buckled-up and buckled-down states (blue and red curves). The sold lines in (a) represent the instantaneous frequency extracted from the signal period, which is in good agreement with the value obtained from the Fourier transform (Fig. 3(h)-(i), symbols). The zero frequency singularity appears as a pronounced dip in the oscillation frequency, which coincides with a jump in the oscillation amplitude. (c) Simulated phase portrait of a resonator ringing-down from a high-amplitude state into the "up" state. The symbols indicate the values at a fixed interval.

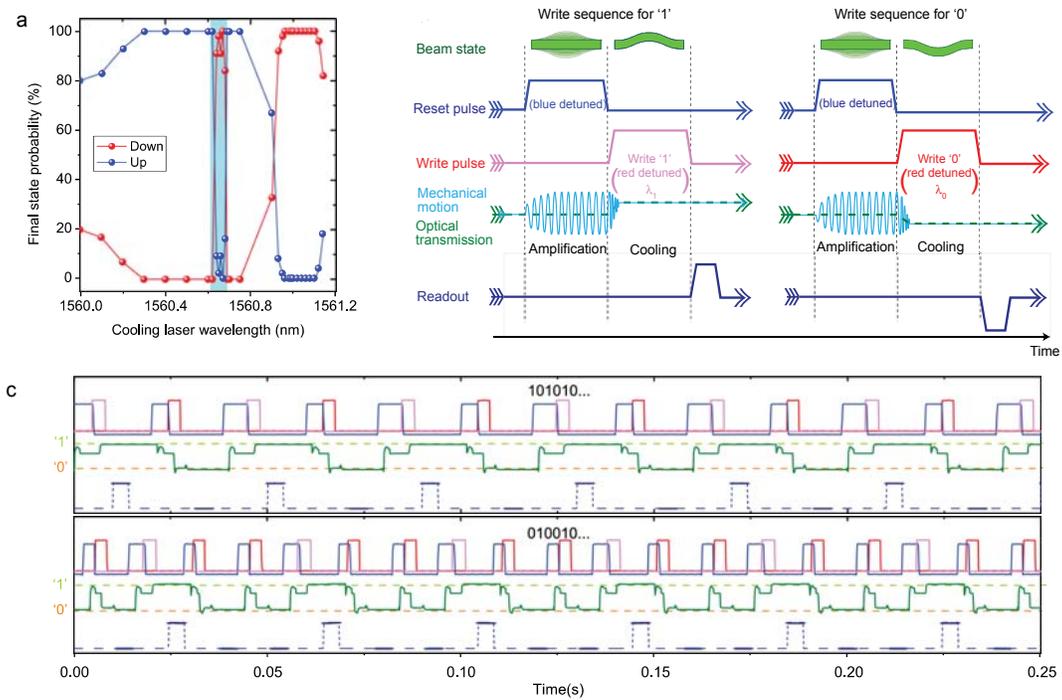

**Figure 5 All-optical, non-volatile nanomechanical memory.** (a) The probability of ending in the up (blue) and down (red) state for different cooling laser wavelengths measured at $P_{in}$= 7.5 dBm. At the wavelengths of the write pulses used for the memory operation there is a 100% probability to end up in the up and down state respectively. Note, that around 1560.6 nm an optomechanical instability occurs (light blue). (b) The sequence employed to implement the memory operation. First, reset pulse (blue) raises the nanomechanical beam above the potential barrier, as indicated by the large oscillations of the mechanical motion (light blue). The subsequent write pulses with wavelength $\lambda_1$ or $\lambda_0$ (pink and red curves) cool the motion of the beam into the bucked-up and buckled-down states, respectively. The dark blue curve represents the weak probe optical signal that monitors the state of the beam. (c) Non-volatile memory operation of the nanomechanical beam. The upper panel shows a series of 10101... written onto the state of the beam, while in the lower panel a series of 01001... bits are stored in the memory element.